\documentclass[peerreview,onecolumn,11pt,draftclsnofoot]{IEEEtran}

\usepackage{amsmath}
\usepackage{amsfonts}
\usepackage{epsfig}
\usepackage{amssymb}
\usepackage{cite}
\usepackage{subfigure}
\usepackage{multirow}
\usepackage{rotating}
\usepackage{graphicx}
\usepackage{tabularx}
\usepackage{array}
\begin{document}

\title{Efficient and Distributed SINR-based Joint Resource Allocation and Base Station Assignment in Wireless CDMA Networks
\thanks{Manuscript received October 2, 2010, revised May 15, 2011. This work was supported in part by Tarbiat Modares University, and in part by Iran Telecommunications Research Center (ITRC) under PhD research grant 89-09-95.}}
\author{Mohammad~R.~Javan and Ahmad~R.~Sharafat, \IEEEmembership{Senior Member, IEEE}
\thanks{The authors are with the Department of Electrical and Computer Engineering, Tarbiat Modares University, P.~O.~Box 14155-4843, Tehran, Iran. Corresponding author is A. R. Sharafat (email: sharafat@modares.ac.ir).}}

\maketitle

\begin{abstract}
We formulate the resource allocation problem for the uplink of
code division multiple access (CDMA) networks using a game theoretic framework, propose an efficient and distributed algorithm for a joint rate and power allocation, and show that the proposed algorithm converges to the unique Nash equilibrium (NE) of the game. Our choice for the utility function enables each user to adapt its transmit power and throughput to its channel. Due to users' selfish behavior, the output of the game (its NE) may not be a desirable one. To avoid such cases, we use pricing to control each user's behavior, and analytically show that similar to the no-pricing case, our pricing-based algorithm converges to the unique NE of the game, at which, each user achieves its target signal-to-interference-plus-noise ratio (SINR). We also extend our distributed resource allocation scheme to multi-cell environments for base station assignment. Simulation results confirm that our algorithm is computationally efficient and its signalling overhead is low. In particular, we will show that in addition to its ability to attain the required QoS of users, our scheme achieves better fairness in allocating resources and can significantly reduce transmit power as compared to existing schemes.
\end{abstract}


\section{Introduction}\label{introduction}
As the demand for high speed services with different quality of
service (QoS) requirements increases, it is becoming more
desirable to improve the efficiency of using the scarce radio
resources. However, temporal variations of system parameters, such
as channels' gains, make this a difficult task. Radio resource
allocation in wireless networks exploits temporal variations in
different users to allocate available resources in such a way that
under some system and service constraints, a performance measure
for each user, e.g., the total throughput or the total transmit
power, is optimized.

In distributed resource allocation, each user determines the
amount of resources to be utilized by that user so that a measure
of QoS can be satisfied. Game theory \cite{fudenberg1} is an
efficient tool that is commonly used in the analysis of
distributed resource allocation schemes. In this context, the
problem of resource allocation is formulated as a non-cooperative
game in which each user chooses a strategy from its strategy
space, such that its utility function is maximized. Distributed
power control has been widely studied in the literature
\cite{foschini1,sung2,mandayam2,shrof2}. In power control, each
user chooses a power level such that its utility function is
maximized in terms of a QoS measure. The utility in \cite{sung2}
is channel capacity, in \cite{mandayam2} is energy efficiency, and
in \cite{shrof2} is a sigmoidal function that approximates the
probability of successful transmission.

Although power control aims to efficiently allocate power levels
to users, if available resources (e.g., power levels, data rates,
and base stations) are jointly allocated, it may be possible to
improve the system's performance even further. However, this comes
at a price. Joint allocation of resources results in
multi-dimensional strategy spaces for users, and requires more
complex utilities, which makes the analysis of the corresponding
game more difficult. Besides, the amount of information needed in
distributed algorithms by each user for joint allocation of
resources is much more than those of single resource allocation
algorithms. Moreover, efficiency of algorithms and their
convergence in joint allocation of resources depend on the utility
function and pricing.

Joint resource allocation has already been addressed in
\cite{hanly1,mandayam1,bedekar1,zander1,zhao1,hayajneh1,
meshkati1, meshkati2, musku1, shrof1}. Joint power allocation and
base station assignment is considered in \cite{hanly1, mandayam1}.
In \cite{hanly1}, the minimum power required to achieve a target
signal-to-interference-plus-noise ratio (SINR) is obtained; and
base stations are assigned in such a way to maintain a minimum
SINR for each user. This approach is satisfactory for voice
service. However, for data services, a more efficient algorithm is
desirable. In \cite{mandayam1} the power control framework
proposed in \cite{mandayam2} is extended for a multi-cell scenario
by modeling the resource allocation problem as a non-cooperative
game in which each user maximizes its own utility, and the
performance of the algorithm depends on the value of pricing.

In \cite{bedekar1, zander1}, joint rate and power allocation for
the downlink of power limited CDMA networks is studied. The
authors in \cite{bedekar1} show that it is optimal for each base
station to transmit to not more than one data-user at any given
time. This scheme is unfair, since users with bad channels or
those that are located farther from the base station would not be
allowed to transmit, as the algorithm only allows the user with
the best channel to transmit. In \cite{zander1}, the authors use
scheduling together with power control to manage interference. In
\cite{zhao1}, joint data rate and power control for CDMA uplink is
obtained by solving the corresponding optimization problem via the
dual Lagrange approach, which needs excessive calculations and
high feedback overhead. In \cite{hayajneh1}, joint data rate and
power control is considered by utilizing a layered game theoretic
approach consisting of two games. In the first game, each user
chooses a rate form its rate strategy space to maximize its
utility function. When this game converges, the target data rate
of each user is determined, and the second game starts. In the
second game, each user chooses a power level to maximize its
utility function. As such, each user plays two consecutive games.
In the first game, there are some parameters that need to be
determined in such a way that the user's data rate is efficient
and achievable. However, \cite{hayajneh1} is silent on how these
parameters are set. Besides, at least the sum of the rates of all
users should be feedbacked in  the first game; and the total
received power should also be feedbacked by the receiver in the
second game.

In \cite{meshkati1,meshkati2}, a joint data rate and power control
scheme that uses the utility proposed in \cite{mandayam2} and QoS
is a function of data rate and queuing delay is considered. The
effects of modulation and constellation size on joint data rate
and power control scheme in \cite{meshkati1} is considered in
\cite{meshkati2}, where the chosen utility function makes it
possible to have many optimal data rates and power levels, i.e.,
multiple Nash equilibria (NEs) exist. In \cite{musku1}, the same
utility function as in \cite{meshkati1,meshkati2} is used and a
joint data rate and power control scheme that may have multiple
NEs is proposed, but no analysis on NEs is provided.

In \cite{shrof1}, joint data rate, power control, and base station
assignment for the downlink of CDMA systems is considered with a
view to maximizing the sum of all users' utilities under total
power constraint, where the optimal data rate is obtained in terms
of the allocated transmit power, i.e., the problem is converted
into a power control one. The optimization problem is solved in
two steps: user selection, and power allocation; and it is shown
that its solution approaches  an optimal one when the number of
users is high. However, it utilizes extensive iterative
signalling, and its base station assignment is sub-optimal.

In this paper, we use game theory \cite{fudenberg1} to formulate
the joint resource allocation problem for the CDMA uplink. Each
user tries to maximize its utility over the data rate and power
level strategy space, independent of other users. Our choice for
the utility function enables each user to efficiently and
adaptively choose its transmit power and data rate. In other
words, to maintain an acceptable QoS, each user increases its
transmit power and/or decreases its throughput when its channel is
not good, and increases its throughput and decreases its transmit
power when its channel is good. In this way, users with bad
channels cause less interference, resulting in higher throughput
values for users with good channels, and in more users supported
by the network.

Due to users' selfish behavior, the output of the game (its NE)
may not be a desirable one, meaning that in such cases, users
transmit at their highest power levels and data rates without
achieving their QoS. To avoid such cases, we use pricing to
control the selfish behavior of users, and propose a distributed
algorithm for joint data rate and transmit power control. We will
show that the proposed game with pricing has a unique NE at which
no user can improve its utility by unilaterally changing its
strategy, and prove that our algorithm converges to the unique NE
of the game.

We further demonstrate that in addition to its distributed nature,
the overhead in our approach is low as compared to other existing
approaches. In particular, we will show that our algorithm needs
the same information as the target SINR-tracking power control
algorithm (TPC) \cite{foschini1} that is known to have a very low
overhead. Note that the strategy space of users in our approach is
two dimensional, which makes the problem more complex as compared
to that of TPC. Nevertheless, we will also show that the proposed
scheme is as computationally efficient as TPC. Moreover, we will
show that in our proposed scheme, each user will achieve its
predefined SINR level at NE. In addition, we will extend our
proposed scheme to multi-cell networks by assigning base stations
to users, and show that for the joint data rate and power level
and base station assignment, there exists a unique NE at which
each user achieves its predefined SINR in multi-cell networks as
well. By way of simulation, we evaluate the performance of our
algorithm for different pricing and QoS levels, and show how the
entry of a new user affects the convergence of the algorithm.

This paper is organized as follows. The problem of joint data rate
and power allocation is formulated in Section \ref{JRPCG}, and the
pricing mechanism and our proposed algorithm are introduced in
Section \ref{JRPCGP}, followed by its extension to a multi-cell
scenario in Section \ref{JRPCGPB}. Practical considerations are
addressed in Section \ref{practical}, simulation results are
presented in Section \ref{simulation}, and conclusions are in
Section \ref{conclusion}.

\section{Non-cooperative Joint Rate and Power Control Game}\label{JRPCG}

In this section, we present the non-cooperative joint data rate
and power control game (NJRPCG) in which pricing is not applied.
Consider a single cell CDMA network with $M$ active users that are
randomly spread in the coverage area. Channel gain from user $i$
to its base station is $g_{i}$, whose bandwidth is $W$. A given
user $i$ transmits at data rate $r_i$ with power $p_i$. SINR of
user $i$ at its base station is \cite{mandayam2,falomari1}
\begin{equation}\label{SINR}
    \gamma_{i} = \frac{W}{r_i}\frac{{g_{i}}{p_i}}{\sum\limits_{j \neq i}{{g_{j}}{p_j}}+N_0},
\end{equation}
where $N_0$ denotes noise power, and $\frac{W}{r_i}$ is the
processing gain. We rewrite (\ref{SINR}) as
\begin{equation}\label{SINR1}
    \gamma_{i} = \frac{W}{r_i}\frac{{p_i}}{R_{i}^\text{eff}},
\end{equation}
where $R_i^\text{eff}$ is the effective interference at the
receiver of user $i$, defined by
\begin{equation}\label{rieff}
    R_{i}^\text{eff}=\frac{\sum\limits_{j \neq i}{{g_{j}}{p_j}}+N_0}{g_{i}}.
\end{equation}
From (\ref{SINR}) and (\ref{rieff}), it is easy to see that the
channel condition for user $i$ depends on the direct channel gain
$g_i$, and on the interference experienced by that user, i.e.,
$\sum\limits_{j \neq i}{{g_{j}}{p_j}}+N_0$. Note that in some
cases, although direct channel gain of a user may be high, the
channel conditions may not be good because of high interference.

For a given modulation type, SINR of a user corresponds to its
received QoS level in terms of its bit error rate (BER). As such,
for a user to achieve its required QoS, it is necessary to
maintain its SINR above a predefined value, i.e.,
$\gamma_i\geq\hat\gamma$. We assume that the transmit power levels
and data rates of users are bounded, i.e., $P_i^\text{min} \leq
{p_i} \leq P_i^\text{max}$, $R_i^\text{min} \leq {r_i} \leq
R_i^\text{max}$.

The NJRPCG game $\mathcal{G}=\langle \mathcal{M},\{
(\mathcal{P}_i,\mathcal{R}_i) \}, \{u_i\} \rangle$ consists of a
set of mobile users $\mathcal{M}=\{1,\cdots,M\}$ that act as
players. The strategy space is $\{(\mathcal{P}_i,
\mathcal{R}_i)\}$, where $\mathcal{P}_i$ is the power strategy set
and $\mathcal{R}_i$ is the data rate strategy set, and $u_i$ is
the utility function for user $i$. The utility of each user is a
function of both transmit power levels and data rates of all
users, i.e., $u_i(\textbf{p},\textbf{r})$. Since the strategy
space of each user is two dimensional, analysis of the game is
more complicated than those of power control games studied in
\cite{sung2,shrof2,mandayam2}. In a NJRPCG, each user maximizes
its own utility in a distributed manner by choosing its transmit
power level and data rate. The objective of the NJRPCG game
$\mathcal{G}$ is
\begin{equation}\label{NJRPCG}
    \max_{(\mathcal{P}_i,\mathcal{R}_i)} ~u_{i}(p_i, \textbf{p}_{-i}, r_i, \textbf{r}_{-i}), \qquad\forall i \in
    \mathcal{M},
\end{equation}
where $\textbf{p}_{-i}$ and $\textbf{r}_{-i}$ are transmit power
and data rate strategies of all users other than user $i$,
respectively. The outcome of the game is the NE at which no user
can increase its utility by unilaterally changing its strategy,
given that other users' strategies are fixed. Formally, NE is
defined as follows.

\emph{Definition 1}: Transmit power level vector $\textbf{p}^*$
and data rate vector $\textbf{r}^*$ is NE of the NJRPCG game
$\mathcal{G}$ if
\begin{equation}\label{NEdefinition}
    ~u_{i}(p_i^*, \textbf{p}_{-i}^*, r_i^*, \textbf{r}_{-i}^*) \geq u_{i}(p_i, \textbf{p}_{-i}^*, r_i, \textbf{r}_{-i}^*), \qquad\forall p_i \in \mathcal{P}_i, \qquad\forall r_i \in \mathcal{R}_i, \qquad\forall i \in \mathcal{M},
\end{equation}

Another definition of NE is based on the best response notion,
which is a set valued function
$\text{br}_i:(\mathcal{P}_{-i},\mathcal{R}_{-i})\rightarrow
(\mathcal{P}_i,\mathcal{R}_i)$ defined by
\begin{equation}\label{bestresponse}
    ~\text{br}_{i}(\textbf{p}_{-i}, \textbf{r}_{-i})=\{p_i \in \mathcal{P}_i, r_i \in \mathcal{R}_i: u_{i}(p_i, \textbf{p}_{-i}, r_i, \textbf{r}_{-i}) \geq u_{i}(\hat p_i, \textbf{p}_{-i}, \hat r_i, \textbf{r}_{-i}), \forall \hat p_i \in \mathcal{P}_i, \hat r_i \in \mathcal{R}_i \}.
\end{equation}
Note that, $\mathcal{P}_{-i}=\prod_{j\neq i} \mathcal{P}_{j}$ and
$\mathcal{R}_{-i}=\prod_{j\neq i} \mathcal{R}_{j}$. As such,
vectors $\textbf{p}^*$ and $\textbf{r}^*$ constitute NE of the
game if and only if for all users $(p_i^*,r_i^*)\in
\text{br}_{i}(\textbf{p}_{-i}^*, \textbf{r}_{-i}^*)$.

In power control, a variety of utility functions can be used. In
\cite{mandayam1}, a utility that corresponds to energy efficiency
is used, i.e., utility is defined as the ratio of the transmitted
information (in bits) per unit of consumed power. In
\cite{shrof2}, a sigmoid like function of SINR is used for
utility. However, for the problem at hand, as stated earlier, due
to the interaction between data rate and transmit power, such
utilities would lead to multiple Nash equilibria.

We choose the utility function in such a way that the following three requirements would be satisfied:
\begin{list}{\labelitemi}{\leftmargin=1em}
\item[1.] Each user aims to achieve higher SINRs. \item[2.] Each
user aims to attain higher data rates. \item[3.] When the channel
is bad and/or when interference is high, each user should increase
its transmit power level and/or decrease its data rate.
\end{list}

Note that the above three requirements may be in conflict with
each other. For example, from (\ref{SINR}), one can easily see
that the above Requirements 1 and 2 are in conflict. In addition,
although each user wants to transmit at higher data rates,
adapting to its channel conditions may decrease its data rate,
meaning that Requirements of 2 and 3 may not be simultaneously
adhered to. As such, we may have to trade-off between the above
requirements. On the other hand, for joint data rate and power
control, the utility is a function of the user's transmit power
and its data rate, and may not be directly a function of SINR.
Besides, in a game theoretic framework, each user aims to maximize
its utility function, and the game will settle at its NE, if one
exits. Hence, each user will achieve an SINR computed by transmit
power levels and data rates of users at the NE. In addition, the
achieved SINR of each user depends on the definition of utility.

We begin by considering the utility of each use as a logarithmic
function of its SINR, as this functions is commonly used for power
control \cite{sung2} and for joint data rate and power control
\cite{zhou1} schemes, i.e.,
\begin{equation}\label{utility1}
    u^1_{i} = \log (1+k \gamma_i),
\end{equation}
where $k$ is an adjustable parameter. Since the utility in
(\ref{utility1}) is an increasing function of SINR, maximizing
this utility in terms of data rate and power level will result in
the user's SINR $\gamma_i$. This is in line with Requirement 1 above. To maximize the utility in (\ref{utility1}),
each user transmits at a high power level and a low data rate,
which contradicts the second point above. To alleviate this, we
add a new term to (\ref{utility1}), which is a logarithmic
function of data rate $r_i$, i.e.,
\begin{equation}\label{utility2}
    u^2_{i} = \log (k' r_i),
\end{equation}
where $k'$ is an adjustable parameter. From (\ref{utility1}) and (\ref{utility2}), the utility of each user is
\begin{equation}\label{utility3}
    u_{i}(p_i, \textbf{p}_{-i}, r_i, \textbf{r}_{-i})= \log (1+k \frac{W}{r_i}\frac{{g_i}{p_i}}{\sum\limits_{j \neq
    i}{{g_j}{p_j}}+N_0})+\log (k' r_i).
\end{equation}
Note that the utility of user $i$ does not depend on data rates of other users, i.e., on $\textbf{r}_{-i}$, but to follow the general rule, we write it in this manner. We rewrite (\ref{utility3}) as
\begin{equation}\label{utility4}
    u_{i}(p_i, \textbf{p}_{-i}, r_i, \textbf{r}_{-i}) = \log (k_1 r_i+k_2 \frac{{p_i}}{R_i^\text{eff}}).
\end{equation}
where $k_1=k'$ and $k_2=k' k W$. In what
follows, we analyze the NJRPCG game (\ref{NJRPCG}) for the utility
function (\ref{utility4}).

\emph{Theorem 1}: There exists a unique NE in the NJRPCG game
(\ref{NJRPCG}).

\emph{Proof}:  The utility function of user $i$, i.e.,
(\ref{utility4}), is an increasing function of $(p_i,r_i)$,
meaning that higher values of transmit power $p_i$ and data rate
$r_i$ lead to a higher utility value. This means that in order to
maximize the utility, each user transmits at its maximum transmit
power and maximum data rate. As such, there exists a unique NE at
which each user transmits at its maximum transmit power and
maximum data rate. ~$\blacksquare$

At NE of (\ref{NJRPCG}), each user transmits at maximum data rate
and maximum power, regardless of channel conditions. However, in general, the data rate should be so chosen to satisfy Shannon capacity. For the game involving (\ref{NJRPCG}), the chosen data rate that maximizes the utility may be above the Shannon capacity. In such cases, satisfying Shannon capacity entails complications that involve both the user and the base station. To avoid such complications, we use pricing for the games in Sections III and IV to improve a user's SINR, which in turn would result in a higher Shannon capacity, meaning that the chosen data rates for those users that with pricing can achieve their target SINRs are feasible, i.e., below Shannon capacity. If the strategy space is bounded, some users may not achieve their target SINRs when pricing is used, and should be removed as their data rates would be above Shannon capacity. Morover, as stated earlier, each user may adapt its transmit power and data rate to its channel conditions (Requirement 3 in Section \ref{JRPCG}). In the next section, we present a pricing mechanism and show that if the strategy space is unbounded, at NE of the game with pricing, each user will achieve its predefined SINR.

\section{Non-Cooperative Joint Rate and Power Control Game with Pricing}\label{JRPCGP}

Transmitting at maximum data rate and maximum power is not always
useful, since the transmit power of each user is considered as
interference to other users. To control the selfish behavior of
each user and adapt its transmit power and/or data rate to the
channel condition, we adjust the parameters of utility function
and use pricing. In doing so, we present a joint data rate and
transmit power control game with pricing (NJRPCGP), show that the
corresponding distributed algorithm converges to the unique NE of
the game, and demonstrate that each user achieves its predefined
SINR. Since the strategy space of users are transmit power levels
and data rates, our pricing depends on both of these values for
each user. The multi-dimensionality of the strategy space of users
means that we have many choices for the pricing function, such as
the weighted sum of users' power levels and data rates, and their
multiplications.

Here, for each users, we use the weighted sum of its squared data
rate and its squared transmit power as the pricing function. Our
choice of the pricing function has the following merits: First,
the utility function remains concave, which is essential for the
game's analysis; and second, the squared function is more
sensitive to variations in its parameters than a linear function.
This means that a user's choice of its strategies highly affects
the value of its pricing factor. We will show in our simulations
that users who transmit at higher power levels and higher data
rates will reduce their transmit power level and data rate more
than those users who transmit at lower power levels and at lower
data rates when pricing is increased. The utility of each user is
\begin{equation}\label{utilitypricing1}
    u_{i} = \log (\alpha_2 r_i+\alpha_1 p_i)- \frac {\lambda}{2}(\frac {\alpha_2}{\alpha_1} r_i^2+\frac {\alpha_1}{\alpha_2}
    p_i^2),
\end{equation}
where $\alpha_1$ and $\alpha_2$ are adjustable parameters and
$\lambda$ is the pricing factor. The utility
(\ref{utilitypricing1}) has the property that for a given value of
the utility function and a fixed $\lambda$, when $\frac
{\alpha_2}{\alpha_1}$ increases, we must decrease the data rate
and/or increase the power level to keep the utility fixed. This
can be used to incorporate Requirement 3 in Section
\ref{JRPCG} into the utility function (\ref{utilitypricing1}). To
this end, we introduce the effective interference $R_i^\text{eff}$
into the utility by
\begin{equation}\label{utilitypricing}
    u_{i} = \log (\alpha_2 R_i^\text{eff} r_i+\alpha_1 p_i)- \frac {\lambda}{2}(\frac {\alpha_2}{\alpha_1} R_i^\text{eff} r_i^2+ \frac {\alpha_1}{\alpha_2} \frac {1}{R_i^\text{eff}} p_i^2).
\end{equation}
In this way, when the effective interference $R_i^\text{eff}$ for
user $i$ increases, that user decreases its data rate and/or
increases its power. We will show that the values of $\alpha_1$
and $\alpha_2$ are important in the actual SINR for each user. For
the above utility, the objective of the NJRPCGP game is
\begin{equation}\label{NJRPCGP}
    \max_{(P_i,R_i)} ~u_{i}(p_i, \textbf{p}_{-i}, r_i, \textbf{r}_{-i}), \forall i \in \mathcal{M}.
\end{equation}

\emph{Theorem 2}: There exists a NE in the above NJRPCGP game.

\emph{Proof}: One can easily show that utility $u_i$ is a jointly
concave function of $(p_i,r_i)$ by forming its second derivatives,
i.e.,
\begin{equation}\label{sdpup}
\frac{\partial^2 u_i}{\partial p_i^2}=-\frac{\alpha_1^2}{(\alpha_1
p_i + \alpha_2 R_i^{\text{eff}} r_i)^2}- \lambda
\frac{\alpha_1}{\alpha_2} \frac{1}{R_i^{\text{eff}}},
\end{equation}
\begin{equation}\label{sdrup}
\frac{\partial^2 u_i}{\partial r_i^2}=-\frac{(\alpha_2
R_i^{\text{eff}})^2}{(\alpha_1 p_i + \alpha_2 R_i^{\text{eff}}
r_i)^2}- \lambda \frac{\alpha_2}{\alpha_2} R_i^{\text{eff}},
\end{equation}
\begin{equation}\label{sdprup}
 \frac{\partial^2 u_i}{\partial p_i \partial r_i}=-\frac{\alpha_1 \alpha_2
R_i^{\text{eff}}}{(\alpha_1 p_i + \alpha_2 R_i^{\text{eff}}
r_i)^2}.
\end{equation}
It is obvious that inequalities in $\frac{\partial^2 u_i}{\partial
p_i^2} \leq 0$, $\frac{\partial^2 u_i}{\partial r_i^2} \leq 0$,
and $\frac{\partial^2 u_i}{\partial p_i^2} \frac{\partial^2
u_i}{\partial r_i^2}- (\frac{\partial^2 u_i}{\partial p_i \partial
r_i})^2 \geq 0$ are strict. Therefore, the utility function is
strictly concave on $(p_i, r_i)$. Besides, the utility function is
continuous in $(\textbf{p},\textbf{r})$. Since the strategy space
$(\mathcal{P}_i,\mathcal{R}_i)$ is a compact, convex, and nonempty
subset of $\mathbb{P}^M \times \mathbb{R}^M$, where $\mathbb{P}^M$
and $\mathbb{R}^M$ are M-dimensional Euclidean space of real
numbers, the proof follows. ~$\blacksquare$

To obtain the best response function for each user $i$, we use the
first derivative of $u_i$ with respect to $(p_i,r_i)$, and write
\begin{equation}\label{derivatives}
\left\{ \begin{array}{ll}
\frac {\partial u_i}{\partial p_i}=0 ~\longrightarrow \frac {\alpha_2 R_i^\text{eff} }{\alpha_1 p_i+ \alpha_2 R_i^\text{eff} r_i } -  \lambda p_i=0,\\
\frac {\partial u_i}{\partial r_i}=0 ~\longrightarrow \frac
{\alpha_1}{\alpha_1 p_i+ \alpha_2 R_i^\text{eff} r_i } - \lambda
r_i=0.
\end{array} \right.
\end{equation}
Rewriting (\ref{derivatives}), we obtain the following system of
equations
\begin{equation}\label{systemequation}
\left\{ \begin{array}{ll}
\alpha_1 \lambda p_i^2 +\alpha_2 \lambda R_i^\text{eff} p_i r_i = \alpha_2 R_i^\text{eff},\\
\alpha_1 \lambda p_i r_i +\alpha_2 \lambda R_i^\text{eff} r_i^2 =
\alpha_1.
\end{array} \right.  
\end{equation}
By solving (\ref{systemequation}), we obtain the following
positive values for $(p_i,r_i)$
\begin{equation}\label{systemsolution}
\left\{ \begin{array}{ll}
p_i =\sqrt{ \frac{1}{2} \frac{\alpha_2}{\alpha_1} \frac{R_i^\text{eff}}{\lambda}},\\
r_i =\sqrt{ \frac{1}{2} \frac{\alpha_1}{\alpha_2} \frac{1}{\lambda
R_i^\text{eff}}}.
\end{array} \right.
\end{equation}

\emph{Remark 1}: From (\ref{systemsolution}), one can observe that
when the channel of user $i$ is bad, i.e., when its channel gain
is low and/or when interference from other users is high, that
user increases its transmit power and/or decreases its data rate.

Using (\ref{systemsolution}), we propose the following iterative
algorithm for updating $(p_i,r_i)$
\begin{equation}\label{update}
   (\textbf{p}^{n+1},\textbf{r}^{n+1})=(\textbf{I}^\text{p}(\textbf{p}^{n},\textbf{r}^{n}),\textbf{I}^\text{r}(\textbf{p}^{n},\textbf{r}^{n})),
\end{equation}
where
\begin{equation}\label{updatefunction}
\left\{ \begin{array}{ll}
I_i^\text{p}(\textbf{p},\textbf{r}) =\sqrt{ \frac{1}{2} \frac{\alpha_2}{\alpha_1} \frac{R_i^\text{eff}}{\lambda}},\\
I_i^\text{r}(\textbf{p},\textbf{r}) =\sqrt{ \frac{1}{2}
\frac{\alpha_1}{\alpha_2} \frac{1}{\lambda R_i^\text{eff}}}.
\end{array} \right.
\end{equation}

Note that if (\ref{update}) converges, it will converge to the
fixed points
$(\textbf{I}^\text{p}(\textbf{p},\textbf{r}),\textbf{I}^\text{r}(\textbf{p},\textbf{r}))$.
Therefore, for the algorithm to converge, it is necessary to have
a fixed point. In general, the fixed point (and the NE of the
game) may not be unique. In such cases, one can devise an
algorithm and derive its initial conditions such that the
algorithm converges to a specific NE as was done in the S-modular
game in \cite{mandayam2} and the joint rate and power control game
in \cite{musku1}. Otherwise, the algorithm may toggle between the
NEs of the game and would not converge. However, if the fixed
point (and the NE) is unique, the algorithm would indeed converge
to its unique fixed point. Now, the remaining questions are under
what conditions, the fixed points exist, and if the fixed point is
unique.

\emph{Remark 2}: Note that both $I_i^\text{p}$ and $I_i^\text{r}$
depend only on $\lambda$, $\alpha_1$, $\alpha_2$, and
$R_i^\text{eff}$, all of which are either locally available or can
be broadcasted by the base station to all users. In other words,
each user $i$ updates its data rate and its transmit power
according to (\ref{updatefunction}) in a distributed manner. In
addition, this means that the feedback overhead of the proposed
scheme is low and are comparable to the information needed by the
well known power control algorithms such as TPC \cite{foschini1}.
Moreover, since each user updates its transmit power and its data
rate according to (\ref{updatefunction}), our scheme needs minimal
calculations.

\emph{Remark 3}: Considering (\ref{updatefunction}), one can
observe that the data rate and the transmit power of a given user
depend on $R_i^\text{eff}$, which can be computed from
(\ref{rieff}). In other words, the data rate of user $i$ does not
have any impact on transmit power levels or on data rates of other
users. Therefore, we can omit the variable $\textbf{r}$ from the
arguments of $\textbf{I}^\text{p}$ and $\textbf{I}^\text{r}$.
Besides, for convergence, it is sufficient to prove the existence
and uniqueness of the fixed point in the transmit power update,
i.e., $\textbf{p}^{n+1}=\textbf{I}^\text{p}(\textbf{p}^{n})$.

From \emph{Remark 3}, we rewrite the transmit power update
function $\textbf{I}^\text{p}(\textbf{p})$ as
\begin{equation}\label{powerupdate}
   I_i^\text{p} (\textbf{p})=\sqrt{ \frac{1}{2} \frac{\alpha_2}{\alpha_1} \frac{1}{\lambda}(\sum\limits_{j \neq i} \frac {g_j}{g_i}{p_j}+ \frac{N_0}{g_i}}).
\end{equation}
Note that when $I_i^\text{p}$ has a unique fixed point, the
iterative transmit power update will globally converge to its
fixed point. To prove the existence of a fixed point, we use the
following theorem.

\emph{Theorem 3 (Brouwer's Fixed Point Theorem)} \cite{border1}:
Let $ \mathcal{\hat P} \subseteq \mathbb{R}^M$ be compact and
convex, and $F:\mathcal{\hat P} \longrightarrow \mathcal{\hat P}$
be a continuous function. There exists a $\textbf{p} \in
\mathcal{\hat P}$ such that $\mathbf{p} =F(\textbf{p})$.

Using Theorem 3, we prove the existence of a fixed point for
$\textbf{I}^\text{p} (\textbf{p})$ in the following theorem.

\emph{Theorem 4}: The function $\textbf{I}^\text{p} (\textbf{p})$
has a fixed point, i.e., there exists a transmit power vector
$\textbf{p}^*$ such that
$\textbf{p}^{*}=\textbf{I}^\text{p}(\textbf{p}^{*})$.

\emph{Proof}: Since the power update function $\textbf{I}^\text{p}
(\textbf{p})$ is continuous, we need to show that there exists a
set $\mathcal{\hat P}\subset \mathbb{R}^M$ such that for each
$\textbf{p} \in \mathcal{\hat P}$, we have $\textbf{I}^\text{p}
(\textbf{p}) \in \mathcal{\hat P}$. From (\ref{powerupdate}), one
can observe that for all $\textbf{p} \geq \textbf{0}$ we have
$I_i^\text{p} (\textbf{p})\geq l_i=\sqrt{ \frac{1}{2}
\frac{\alpha_2}{\alpha_1} \frac{1}{\lambda}(\frac{N_0}{g_i}})$. We
define $\underline{l}= \min_{i} l_i$, $c_j=\max_{i} \frac{1}{2}
\frac{\alpha_2}{\alpha_1} \frac{1}{\lambda}\frac{g_j}{g_i}$, and
$\overline{c}=\max(\max_i c_i, \max_i l_i)$. As such,
$x^2=(M-1)\overline{c}x+\overline{c}$ is well defined. This
equation always has two roots, and the larger one is denoted by
$\overline{x}$. We define the set $\mathcal{\hat P}=
\{\textbf{p}:\underline{l} \leq p_i\leq \overline{z}\}$ where
$\overline{z} > \overline{x}$. Note that for each $\textbf{p} \in
\mathcal{\hat P}$, we have $\textbf{I}^\text{p} (\textbf{p}) \in
\mathcal{\hat P}$. Therefore, we have a continuous function
$\textbf{I}^\text{p} (\textbf{p})$ such that $\textbf{I}^\text{p}:
\mathcal{\hat P} \longrightarrow \mathcal{\hat P}$. From Theorem
3, the function $\textbf{I}^p$ has a fixed point. ~$\blacksquare$

\emph{Remark 4}: From Theorem 4, it follows that the power update
function $\textbf{I}^\text{p}(\textbf{p})$ always has a fixed
point, and if it is unique, it globally converges to the unique
fixed point. This is in contrast to TPC \cite{foschini1} that only
under some conditions has a fixed point \cite{zander2}. Moreover,
this means that we do not need the compactness assumption as in
Theorem 2, i.e., the strategy space of users can be unbounded.
This property does not hold in \cite{meshkati1,meshkati2, musku1}.

In addition to the existence of a fixed point, in Theorem 5, we
prove that the fixed point is unique, which guarantees that the
algorithm globally converges to this unique NE.

\emph{Theorem 5}: The fixed point of $\textbf{I}^\text{p}
(\textbf{p})$ is unique.

\emph{Proof}: We use the notion of standard functions defined in
\cite{yates2}. A standard function $\textbf{I}(\textbf{p})$ has
the following three properties for all $\textbf{p} \geq
\textbf{0}$:

1. Positivity: $\textbf{I}(\textbf{p})> \textbf{0}$

2. Monotonicity: if $ \textbf{p} > \textbf{p}' $, then
$\textbf{I}(\textbf{p})\geq \textbf{I}(\textbf{p}')$

3. Scalability: for all $a>1$, $a \textbf{I}(\textbf{p})\geq
\textbf{I}(a \textbf{p})$

In Theorem 1 in \cite{yates2}, it is shown that when a standard
function has a fixed point, that fixed point is unique. One can
easily show that $\textbf{I}^\text{p}(\textbf{p})$ is a standard
function. From Theorem 4, we know that
$\textbf{I}^\text{p}(\textbf{p})$ has a fixed point, and thus, the
fixed point is unique. ~$\blacksquare$

From Theorems 4 and 5 together with Remark 3, it follows that the
iterative updating function in (\ref{update}) converges to a
unique fixed point. Therefore, the output of the joint data rate
and transmit power control game with pricing is a unique NE, and
the distributed data rate and transmit power updates converge to
this unique NE. In what follows, we prove an interesting property
of NE of NJRPCGP game in (\ref{NJRPCGP}).

\emph{Theorem 6}: At NE, data rate and transmit power of each user
are related via
\begin{equation}\label{nashproperty}
    \frac{1}{r_i} \frac{p_i}{R_i^\text{eff}}=
    \frac{\alpha_2}{\alpha_1}.
\end{equation}

\emph{Proof}: The proof follows from (\ref{systemsolution}).
~$\blacksquare$

Theorem 6 has the following interesting notion. We rewrite
(\ref{SINR1}) as
\begin{equation}\label{SINR2}
    \gamma_{i} =
    \frac{W}{r_i}\frac{{p_i}}{R_i^\text{eff}}=\frac{\alpha_2}{\alpha_1}W.
\end{equation}
Therefore, when the required SINR for user $i$ is $\gamma_i$ in
(\ref{SINR2}), that user will attain its SINR at NE. We can set
different values of $\alpha_1$ and $\alpha_2$ for different users,
resulting in different values of SINR. Note that the SINR value
$\frac{\alpha_2}{\alpha_1}W$ is the SINR that any user can achieve
by maximizing its utility. In Theorem 6, this SINR is predefined
as $\frac{\alpha_2}{\alpha_1}W$, which can be set by the base
station and/or the service provider based on the service and the
user's quality of service.

In the above, we assumed that the convergence point of the
iterative updating function (\ref{updatefunction}) is in the
strategy space of each user, and showed that under this
assumption, the distributed joint data rate and transmit power
allocation algorithm always converges to a point at which each
user achieves its target SINR. But, this assumption may not be
valid in general. In reality, bounds on users' data rates and
power levels as well as channel conditions can move the
convergence point of the iterative updating function
(\ref{updatefunction}) outside the strategy space defined for each
user. In what follows, we analyze this case.

Consider the utility and the game formulation in
(\ref{utilitypricing}) and (\ref{NJRPCGP}), respectively, and the
strategy space of each user as in Section \ref{JRPCG}. When the
convergence point of the iterative updating function
(\ref{updatefunction}), i.e., $(p_i,r_i)$ is such that
$P_i^\text{min} \leq {p_i} \leq P_i^\text{max}$, $R_i^\text{min}
\leq {r_i} \leq R_i^\text{max}$, the solution of the game is
(\ref{systemsolution}) and the iterative updating is
(\ref{update}). When data rate of a user reaches its upper or
lower bounds, the solution for data rate is that bound (i.e.,
$r_i=R_i^\text{bound}$ where $R_i^\text{bound}=R_i^\text{min}$ if
it reaches the lower bound, and $R_i^\text{bound}=R_i^\text{max}$
if it reaches the upper bound). To obtain the power level, in
(\ref{utilitypricing}) we replace $r_i$ with $R_i^\text{bound}$,
take the derivative with respect to $p_i$, and write
\begin{equation}\label{powerequation}
    \alpha_1 \lambda p_i^2 +\alpha_2 \lambda R_i^\text{eff} R_i^\text{bound}
    p_i-\alpha_2 R_i^\text{eff}=0.
\end{equation}
The iterative transmit power update function $\textbf{I}^\text{p}$
is obtained from the positive root of (\ref{powerequation}) as
\begin{equation}\label{powerupdater}
    I_i^\text{p}(\textbf{p}) = \frac{-\alpha_2 \lambda R_i^\text{eff} R_i^\text{bound}+\sqrt{(\alpha_2 \lambda R_i^\text{eff} R_i^\text{bound})^2+
    4 \alpha_1 \alpha_2 \lambda R_i^\text{eff}}}{2 \alpha_1 \lambda}.
\end{equation}
The power level of a user is bounded by its upper or lower limits.
When transmit power of a user reaches its upper or lower bounds,
to obtain the data rate,  we replace $p_i$ with $P_i^\text{bound}$
in (\ref{utilitypricing}), take its derivative with respect to
$r_i$, and write
\begin{equation}\label{rateequation}
    \alpha_2 \lambda R_i^\text{eff} r_i^2+ \alpha_1 \lambda P_i^\text{bound}
    r_i-\alpha_1=0.
\end{equation}
The iterative data rate update function $\textbf{I}^\text{r}$, is
obtained from the positive root of (\ref{rateequation}) as
\begin{equation}\label{rateupdater}
    I_i^\text{r}(\textbf{p}) = \frac{-\alpha_1 \lambda P_i^\text{bound}+\sqrt{(\alpha_1 \lambda P_i^\text{bound})^2+ 4 \alpha_1 \alpha_2 \lambda R_i^\text{eff}}}{2 \alpha_2 \lambda
    R_i^\text{eff}}.
\end{equation}

Since (\ref{powerupdater}) is a standard function, one can easily
prove the following theorem.

\emph{Theorem 7}: Consider the NJRPCGP game (\ref{NJRPCGP}). The
distributed joint data rate and transmit power control algorithm
will always converge to the unique NE of the game.

When the convergence point of the iterative updating function
(\ref{updatefunction}), i.e., $(p_i,r_i)$ is within the strategy
space of users, all users will achieve their target SINRs. This is
not true when this point is outside the strategy space of users,
resulting in some users reaching their lower bound and/or upper
bound on data rate and/or transmit power. Note that a user
achieves a higher SINR by increasing its transmit power and/or by
decreasing its data rate. Therefore, the achieved SINR of those
users who reach their lower bound on data rate and/or upper bound
on transmit power is below their target SINR. Such users consume
network resources without achieving their required QoS, and cause
interference to other users as well. To deal with such cases,
depending on service provisioning policies by the network
provider, either these users should be removed from the network
\cite{rasti1, rasti2}, or pricing should be changed
\cite{mandayam3}. Interference caused by those users that cannot
attain their target SINRs can be reduced by removing them from the
network one by one until all remaining users achieve their target
SINR. Interference caused by those users whose achieved SINRs is
higher than their target SINRs can be reduced by forcing such
users to reduce their transmit power levels and/or data rates.

The non-cooperative joint data rate and power control game with
pricing is stated below.

~~~ \emph{\textbf{NJRPCGP Algorithm}}:
\begin{list}{\labelitemi}{\leftmargin=3em}
\item[1.] each user begins transmitting at power level $p_i(t_0)$
and data rate $r_i(t_0)$ constituting the power vector
$\textbf{p}(t_0)$ and the data rate vector $\textbf{r}(t_0)$.
\item[2.] At each iteration time $t_k$, each user updates its the
transmit power level and its data rate according to
(\ref{updatefunction}):
\begin{list}{\labelitemi}{\leftmargin=3em}
\item[a.] When $P_i^\text{min} \leq {p_i} \leq P_i^\text{max}$ and
$R_i^\text{min} \leq {r_i} \leq R_i^\text{max}$, we set
$p_i(t_k)=p_i$ and $r_i(t_k)=r_i$. \item[b.] When $P_i^\text{min}
\leq {p_i} \leq P_i^\text{max}$, and ${r_i} \leq R_i^\text{min}$
or $R_i^\text{max}\leq {r_i}$, we update $p_i(t_k)$ from
(\ref{powerupdater}) and set $r_i(t_k)=R_i^\text{min}$ or
$r_i(t_k)=R_i^\text{max}$, respectively. \item[c.] When ${p_i}
\leq P_i^\text{min}$ or $P_i^\text{max} \leq {p_i}$ and
$R_i^\text{min} \leq {r_i} \leq R_i^\text{max}$, we set
$p_i(t_k)=P_i^\text{min}$ or $p_i(t_k)=P_i^\text{max}$,
respectively, and update $r_i(t_k)$ from (\ref{rateupdater}).
\end{list}
\item[3.] When $\max_i \left(
\|p_i(t_k)-p_i(t_{k-1})\|+\|r_i(t_k)-r_i(t_{k-1})\| \right) \leq
\delta$, we stop; else we go to Step 2 above.
\end{list}

\section{Base Station Assignment}\label{JRPCGPB}

In this section, we extend our proposed scheme to a multi-cell
network in which active users communicate through $B$ base
stations. The base station assigned to user $i$ is $a_i$, and
$g_{a,i}$ denotes the channel gain from the transmitter of user
$i$ to base station $a$. Each user transmits at data rate $r_i$ at
power $p_i$. The SINR of user $i$ at its base station $a_i$ is
\begin{equation}\label{SINRmulticell}
    \gamma_{a_i,i} = \frac{W}{r_i}\frac{{g_{a_i,i}}{p_i}}{\sum\limits_{j \neq i}{{g_{a_i,j}}{p_j}}+N_0},
\end{equation}
where we assume noise power $N_0$ is the same for all base
stations.

In a multi-cell network, base station assignment must also be
considered. We incorporate base station assignment into our joint
data rate and transmit power level scheme proposed in Section
\ref{JRPCGP}, i.e., each user chooses its base station in a
distributed manner in addition to setting its transmit power level
and data rate. Since utility of each user is a decreasing function
of its effective interference $R_i^\text{eff}$, we use the value
of $R_i^\text{eff}$ as a measure for base station assignment. The
objective of the non-cooperative joint data rate, transmit power
level, and base station assignment game with pricing (NJRPCGPB) is
\begin{equation}\label{NJRPCGPB}
   \max_{(P_i,R_i)} ~u_{a_i,i}(p_i, \textbf{p}_{-i}, r_i, \textbf{r}_{-i}), \forall i \in \mathcal{M},
\end{equation}
where
\begin{equation}\label{basestation}
   a_i=\text{argmin}_{a} ~R_{a,i}^\text{eff},
\end{equation}
and
\begin{equation}\label{utilitybasestation}
     u_{a,i} = \log (\alpha_2 R_{a,i}^\text{eff} r_i+\alpha_1 p_i)- \frac {\lambda}{2}(\frac {\alpha_2}{\alpha_1} R_{a,i}^\text{eff} r_i^2+
    \frac {\alpha_1}{\alpha_2} \frac {1}{R_{a,i}^\text{eff}} p_i^2),
\end{equation}
where $R_{a,i}^\text{eff}$ is defined in (\ref{rieff}). Dynamic
base station assignment according to (\ref{basestation}) means
that $R_{a_i,i}^\text{eff} \leq R_{a,i}^\text{eff}$ is satisfied
for all base station assignments. Inverting this inequality and
multiplying both sides by $\frac{p_i}{r_i}$, we get
\begin{equation}\label{maxsinrallocation}
     \frac{p_i}{r_i} \frac{1}{R_{a_i,i}^\text{eff}} \geq \frac{p_i}{r_i}
     \frac{1}{R_{a,i}^\text{eff}}.
\end{equation}
Equation (\ref{maxsinrallocation}) has the following
interpretation. For a fixed value of $\frac{p_i}{r_i}$, each user
is assigned to a base station in which its SINR is the highest
compared to all other choices for base station. When target SINR
is predefined via the values of $\alpha_1$, $\alpha_2$ and
$\lambda$ before the game starts, each user is assigned to a base
station for which it transmits at a lower power level and a higher
data rate as compared to all other choices for base station.

Note that for each value of $R_{a,i}^\text{eff}$, the utility
function of each user is a jointly concave function of
$(p_i,r_i)$, and hence $u_{a_i,i}$ is a concave function, and the
game (\ref{NJRPCGPB}) is well defined. From (\ref{basestation}),
(\ref{utilitybasestation}), and (\ref{updatefunction}), one can
derive the power and data rate update functions when base station
is assigned. We propose the following iterative algorithm for
updating $(p_i,r_i)$:
\begin{equation}\label{updatebasestation}
   (\textbf{p}^{n+1},\textbf{r}^{n+1})=(\textbf{I}^\text{p}(\textbf{p}^{n},\textbf{r}^{n}),\textbf{I}^\text{r}(\textbf{p}^{n},\textbf{r}^{n})),
\end{equation}
where
\begin{equation}\label{updatefunctionbasestation}
\left\{ \begin{array}{ll}
I_i^\text{p}(\textbf{p},\textbf{r}) =\min_{a} I_{a,i}^\text{p}(\textbf{p},\textbf{r}),\\
I_i^\text{r}(\textbf{p},\textbf{r}) =\max_{a}
I_{a,i}^\text{r}(\textbf{p},\textbf{r}),
\end{array} \right.
\end{equation}
and
\begin{equation}\label{userupdatefunctionbasestation}
\left\{ \begin{array}{ll}
I_{a,i}^\text{p}(\textbf{p},\textbf{r}) =\sqrt{ \frac{1}{2} \frac{\alpha_2}{\alpha_1} \frac{R_{a,i}^\text{eff}}{\lambda}},\\
I_{a,i}^\text{r}(\textbf{p},\textbf{r}) =\sqrt{ \frac{1}{2}
\frac{\alpha_1}{\alpha_2} \frac{1}{\lambda R_{a,i}^\text{eff}}}.
\end{array} \right.
\end{equation}

From Theorem 5 in \cite{yates2}, the power update function in
(\ref{updatefunctionbasestation}) is a standard function. Hence
all results in Section \ref{simulation} for the game
(\ref{NJRPCGP}) also hold for the game (\ref{NJRPCGPB}). In
particular, we have the following theorem.

\emph{Theorem 8}: The game (\ref{NJRPCGPB}) has a unique NE to
which our distributed algorithm (\ref{updatebasestation})
converges. When the convergence point of the iterative updating
function is within the strategy space of users, each user will
achieve its target SINR.

The non-cooperative joint rate, power, and base station assignment
game with pricing is stated below.

~~~ \emph{\textbf{NJRPCGPB algorithm}}:
\begin{list}{\labelitemi}{\leftmargin=3em}
\item[1.] each user connects randomly to one base station and
begins transmitting at the power level $p_i(t_0)$ and the data
rate $r_i(t_0)$ constituting the rate vector $\textbf{r}(t_0)$ and
power vector $\textbf{p}(t_0)$. \item[2.] At each iteration time
$t_k$, each user updates its base station to which it connects,
i.e., $a_i$, according to (\ref{basestation}). \item[3.] Each user
updates the transmit power level and rate according to
(\ref{userupdatefunctionbasestation}):
\begin{list}{\labelitemi}{\leftmargin=3em}
\item[a.] When $P_i^\text{min} \leq {p_i} \leq P_i^\text{max}$ and
$R_i^\text{min} \leq {r_i} \leq R_i^\text{max}$, we set
$p_i(t_k)=p_i$ and $r_i(t_k)=r_i$. \item[b.] When $P_i^\text{min}
\leq {p_i} \leq P_i^\text{max}$, and ${r_i} \leq R_i^\text{min}$
or $R_i^\text{max}\leq {r_i}$, we update $p_i(t_k)$ from
(\ref{powerupdater}) and set $r_i(t_k)=R_i^\text{min}$ or
$r_i(t_k)=R_i^\text{max}$, respectively. \item[c.] When ${p_i}
\leq P_i^\text{min}$ or $P_i^\text{max} \leq {p_i}$ and
$R_i^\text{min} \leq {r_i} \leq R_i^\text{max}$, we set
$p_i(t_k)=P_i^\text{min}$ or $p_i(t_k)=P_i^\text{max}$,
respectively, and update $r_i(t_k)$ from (\ref{rateupdater}).
\end{list}
\item[4.] When $ \max_i
\left(\|p_i(t_k)-p_i(t_{k-1})\|+\|r_i(t_k)-r_i(t_{k-1})\| \right)
\leq \delta$, we stop; else we go to Step 2 above.
\end{list}

Note that the NJRPCGPB algorithm is similar to the NJRPCGP
algorithm except in Step 2, where base stations are assigned to
users.

\section{Practical Considerations}\label{practical}

In this section, we explain some practical issues concerning
system parameters, message passing, and discrete data rates in the
proposed algorithms.

\subsection{System Parameters}\label{systemparameters}

There are three parameters in the utility function, namely
$\alpha_1$, $\alpha_2$, and $\lambda$; each with a different
impact on data rates and power levels of users. In Theorem 6, the
impacts of $\alpha_1$ and $\alpha_2$ are shown. As stated in
(\ref{SINR2}), at the NE of the game (games in Sections
\ref{JRPCGP} and \ref{JRPCGPB}), depending on the values of
$\alpha_1$ and $\alpha_2$, each user will achieve a specific SINR.
This means that by setting the values of $\alpha_1$ and
$\alpha_2$, the target SINR for each user is set. In general, the
values of $\alpha_1$ and $\alpha_2$ can be set by the base
stations and/or by the service provider based on the service and
the user's required quality of service. As such, depending on
channel conditions, each users chooses its transmit power and data
rate to achieve its SINR. As an example, suppose that in a network
with $W=10^6$ Hz bandwidth, the required quality of service for
user $i$ in terms of its SINR is $\gamma_i=20$. Considering
(\ref{SINR2}), we have
$\frac{\alpha_2}{\alpha_1}=\frac{\gamma_i}{W}=2\times 10^{-5}$. In
this case, all values of $\alpha_1$ and $\alpha_2$ that satisfy
$\frac{\alpha_2}{\alpha_1}=2 \times 10^{-5}$, e.g.,
$\alpha_1=10^6$ and $\alpha_2=20$, or $\alpha_1=15$ and
$\alpha_2=3\times 10^{-4}$ are acceptable. This means that only
the ratio $\frac{\alpha_2}{\alpha_1}$ is important, rather than
the values of $\alpha_1$ and $\alpha_2$.

The pricing $\lambda$ can be different for each user, and affects
its transmit power and data rate. Increasing $\lambda$ would
decrease the transmit power and the data rate, and decreasing
$\lambda$ would result in the opposite. Depending on channel
conditions and strategy bounds, the convergence point of the
algorithm may not be within the strategy space, meaning that at
NE, some users may not achieve their predefined SINR. By
increasing $\lambda$, some users decrease their transmit power,
and it may possible that more users achieve their target SINR.
However, finding the optimal value of $\lambda$ is not easy, and
in many cases a heuristic approach is needed to tune the pricing.

Depending on problem formulation and objectives, one can include
many factors into pricing. As the number of users increases,
interference also increases. This is undesirable, as it may cause
some users not to achieve their SINRs. However, one can include
the number of users in the pricing, i.e., $\lambda_i=c M$, so that
when the number of users increases, pricing would increase as
well. The direct path gain of a user to its receiver is also an
option. By proper inclusion of this parameter into pricing, one
can achieve better results. For example, if pricing is an
increasing function of the direct channel gain $\lambda_i=c g_i$,
the algorithm favors those users whose channel gains are low. In
this way, one can improve fairness. On the other hand, if pricing
is a decreasing function of the direct channel gain
$\lambda_i=\frac{c}{g_i}$, the algorithm favors those users whose
channel gains are high. In our case that we wish to achieve
predefined SINRs at NE, we include the target SINR into pricing
similar to the case of direct channel gains. This would make the
algorithm favorable to users with low SINR or high SINR, i.e.,
$\lambda_i=c \frac{\alpha_2}{\alpha_1}$ or $\lambda_i=c
\frac{\alpha_1}{\alpha_2}$, respectively.

For the analysis in Section \ref{JRPCGPB}, we cannot choose all of
the above mentioned values for $\lambda_i$, as the iterative power
update function in (\ref{updatefunctionbasestation}) may not be a
standard function. To guarantee that the power update function is
a standard function, the value of $\lambda_i$ must be the same for
all base stations. This means that the choice of $\lambda_i=c
g_{a,i}$ or $\lambda_i=\frac{c}{g_{a,i}}$ may prevent the power
update function to be a standard one, and could not be used, since
the value of $g_{a,i}$ may be different for each base station $a$.

Choosing the value of $c$ is not easy, and one may use heuristics
to choose its value. In our case, the achievable SINR at NE is
predefined, and based on channel conditions and strategy bounds,
the convergence point of the algorithm may be out of the strategy
space. In such cases, users are divided into 3 categories: users
that achieve SINRs below their target SINRs, users that achieve
their target SINRs, and users that achieve SINRs above their
target SINRs. In our case, only users in the first category do not
achieve their required QoS. As pricing increases, the strategy
chosen by users tends towards the lower bound in the strategy
space; and when pricing decreases, users' strategy tends towards
the upper bound in the strategy space. An increase in pricing
makes it more likely that all users achieve their target SINRs.
Hence, when there are some category 1 users, pricing should be
increased by the value of $\Delta c$. This should be repeated
until all users achieve their target SINRs. In this way, the least
value of pricing at which all users achieve their target SINR is
selected, meaning that users are allowed to utilize more resources
so long as all users can achieve their target SINRs. Note that
further increases in pricing will allow entry of new users, as
network resources are not completely utilized. However, when
increased pricing is not effective, some users may have to be
removed.

\subsection{Message Passing}\label{feedbackissues}
In our proposed algorithm, for each user to update its transmit
power and data rate, it needs to know the value of
$R_i^{\text{eff}}$, which can be obtained from the value of direct
channel gain $g_i$ and the total received power from all users
plus noise power at the base station, i.e.,
$\sum\limits_{j}{{g_{j}}{p_j}}+N_0$. It is assumed that the base
station broadcasts the value of
$\sum\limits_{j}{{g_{j}}{p_j}}+N_0$ with fixed power. For the
value of $g_i$, either the base station can transmit it to users
or the user can estimate it. In the first case, when the number of
users is high, the amount of message passing would be high.
Estimating the channel gains by users would be simple in the
frequency-division duplexed systems, or in time-division duplexed
systems. In such cases, users calculate the interference caused to
them from the direct channel gain and the total interference plus
noise power at the base station, i.e., $\sum\limits_{j \neq
i}{{g_{j}}{p_j}}+N_0$, by subtracting ${g_{i}}{p_i}$ from
$\sum\limits_{j}{{g_{j}}{p_j}}+N_0$. Note that with the same
amount of message passing as for the TPC, we allocate the transmit
power as well as the data rate. The amount of information is also
the same as in \cite{musku1}, but as we will show in Section
\ref{simulation}, the performance of our algorithm is better than
that of \cite{musku1}, as our algorithm consumes much less power
than \cite{musku1}.

\subsection{Discrete  Data Rates}\label{discreterates}
In our framework and proposed algorithm, we assumed that the data
rate for each user is a continuous variable, which may not be true
in some actual modulation schemes. In most cases, an algorithm
developed for continuous data rates cannot be applied to discrete
data rates, as the continuous assumption is crucial for them. In
the game theoretic framework in this paper, each user aims to
maximize its utility over a continuous domain of strategy space.
In such cases, it may be possible to use the nearest data rate
from the discrete data rate values. However, in general, one
cannot substitute a continuous variable with a discrete one. This
is because of the coupling between the data rate and other
variables in the system. In such cases, the algorithm may toggle
between two points.

In our proposed scheme, as can be seen from (\ref{updatefunction})
and Remark 3, the data rate of each user has no effect on its
transmit power. Instead, only the transmit power levels of users
are coupled, and convergence of the algorithm depends only on
convergence of the power update function. This would make it
possible to consider discrete data rates. To do this, each user
updates its transmit power and data rate according to
(\ref{updatefunction}), and then chooses the nearest lower data
rate from the set of discrete rates available to this user. This
would guarantee that the achieved SINR is above the target SINR,
as evidenced from (\ref{nashproperty}) and (\ref{SINR2}).

\section{Simulation Results}\label{simulation}
We now present simulation results for our distributed joint data
rate and power control algorithm. The system under study is the
uplink of a CDMA network consisting of two base stations and 5
users located at [110, 130, 210, 390, 410] meters from the first
base station and $[410, 390, 310, 130, 110]$ meters from the
second base station. The channel gain from user $i$ to base
station $a$ is as $g_{a,i}= \frac {\xi}{d_{a,i}^\eta}$, where
$d_{a,i}$ is the distance between user $i$ and base station $a$,
$\eta$ is the path loss exponent, and $\xi$ models power
variations due to shadowing. We set $\eta=4$ and $\xi = 0.097$ as
in \cite{musku1}. The bandwidth is $W=10^6$ Hz.

\begin{figure}[t]
  \begin{center}
    \includegraphics[width=9 cm , height=6 cm]{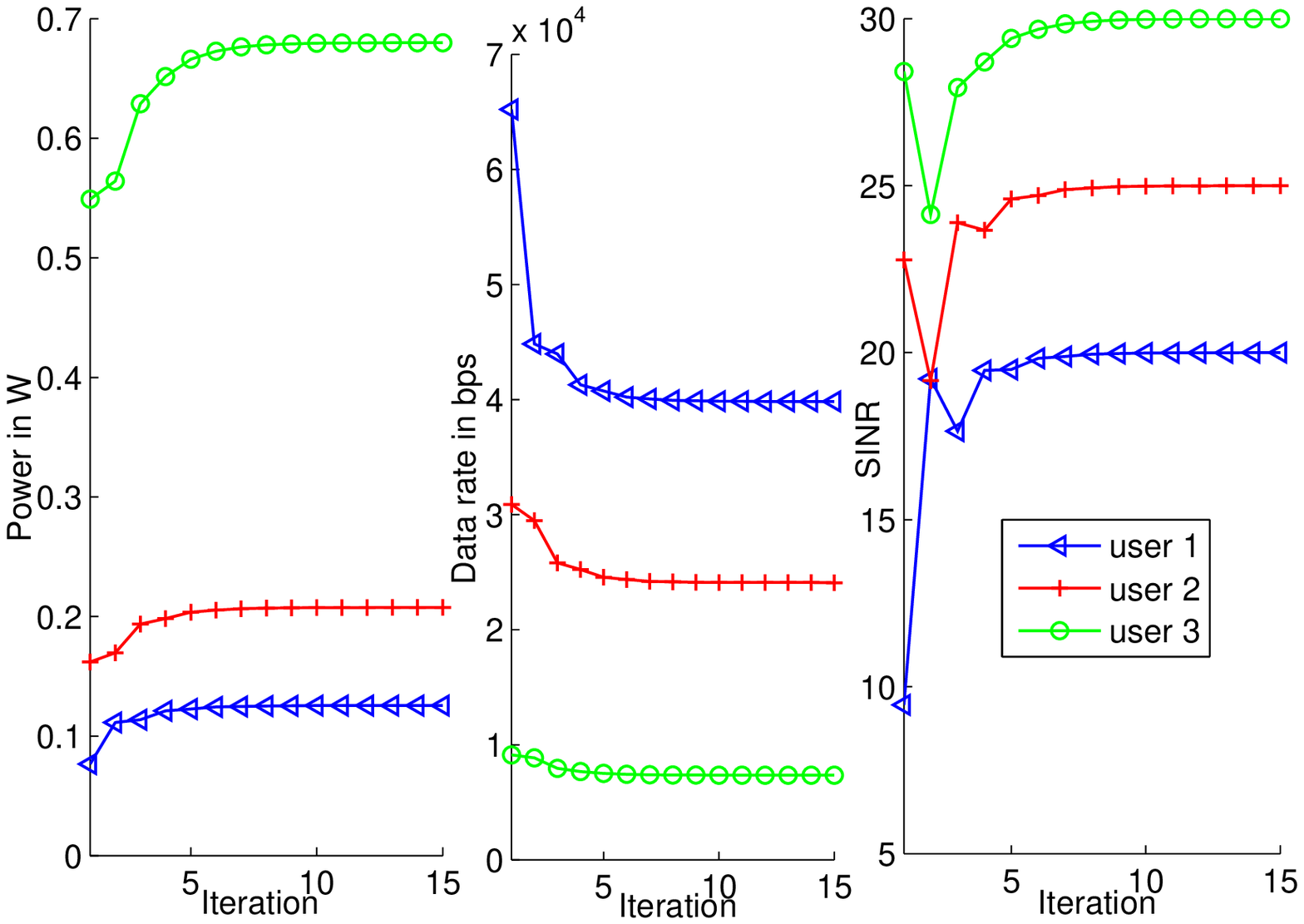} 
    \caption{Transmit power levels, data rates, and attained SINRs for our proposed resource allocation algorithm for different target SINRs. Note that every user attains its target SINR.} 
    \label{fig:Power_rate_SINR_dif_SINR}
  \end{center}
\end{figure}

Consider a network consisting of the first base station and the
first three users. We obtain transmit power levels and data rate
updates (\ref{update}) for $\alpha_1=10^6$ for all users. We set
different values for $\alpha_2$ for different users, i.e.,
$\alpha_2=[20, 25, 30]$ for users 1, 2, and 3, respectively.
Pricing factor is set to $\lambda=10^{-4}$.
Fig.~\ref{fig:Power_rate_SINR_dif_SINR} shows that every user
attains its target SINR.

\begin{figure}[t]
  \begin{center}
    \includegraphics[width=9 cm , height=6 cm]{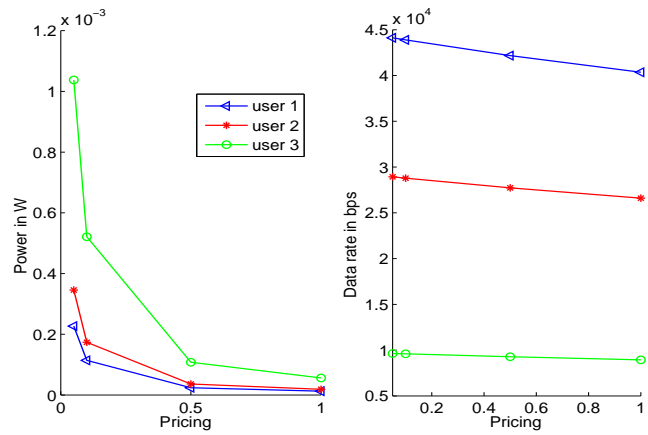} 
    \caption{The impact of increased pricing on the converged values of transmit power levels and data rates.} 
    \label{fig:Power_rate_SINR_price_variable}
  \end{center}
\end{figure}

Next, we consider pricing in the range $0.05$ to $1$ for each
user, and set $\alpha_1=10^6$ and $\alpha_2=20$ for all users so
that their target SINRs are equal to 20. Fig.
\ref{fig:Power_rate_SINR_price_variable} shows that as pricing
increases, each user decreases its data rate and transmit power.
In addition, as pricing increases, users that transmit with more
power and higher data rates decrease their transmit power and data
rate more than those whose pricing is less. This, as stated
before, is because of the pricing function which is a squared
function of transmit power and data rate. Note that all users
achieve their target SINRs regardless of the value of pricing.

We also show the impact of bounding the strategy space (data rates
and transmit power levels) of users. Here, pricing is
$\lambda=10^{-5}$, and $\alpha_1=10^6$, $\alpha_2=20$ for all
users. Table \ref{tableboundingstrategy} shows that user 2
achieves its target SINR by adjusting its transmit power and data
rate. The achieved SINR by user 1 who enjoys a good channel is
above its target SINR; whereas user 3 whose channel is bad does
not attain its target SINR. When we increase pricing from
$\lambda=10^{-5}$ to $\lambda=10^{-4}$, all users achieve their
target SINRs by appropriate reductions in their transmit power
levels and data rates. Next, We set the pricing back to
$\lambda=10^{-5}$ and remove user 3 from the network. Note that
the remaining users increase their data rates and reduce their
transmit power levels, and the achieved SINR of user 1 is above
its target SINR.

\begin{figure}[t]
  \begin{center}
    \includegraphics[width=9 cm , height=6 cm]{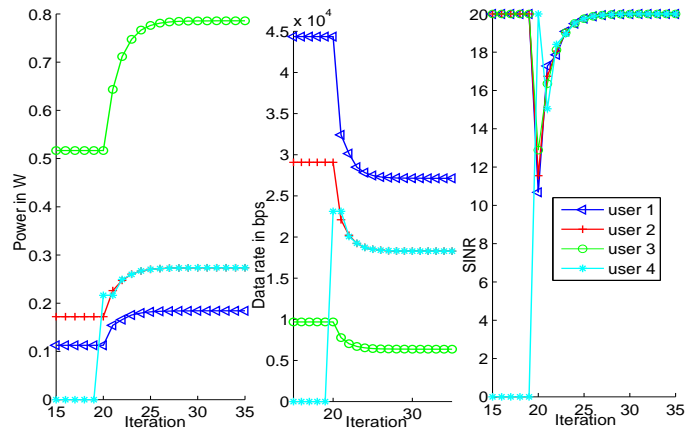} 
    \caption{The impact of a new user on transmit power levels, data rates, and attained SINR in our proposed resource allocation algorithm. Note that a new user only temporarily affects the attained SINR of existing users.} 
    \label{fig:Power_rate_SINR_new_user}
  \end{center}
\end{figure}

We now show what happens when a new user enters the system. The
setup is similar to the one for the first example above, but after
a time period, a new user (user $4$) enters the system. This user
is located 130 meters from the base station (same as user 2).
Here, pricing is $\lambda=10^{-4}$. Fig.
\ref{fig:Power_rate_SINR_new_user} shows iterations 15 to 35 of
our algorithm, where the new user enters at iteration 20. As shown
in Fig. \ref{fig:Power_rate_SINR_new_user}, when user 4 starts its
transmission, it increases interference, i.e., reduces the actual
SINR of other users. But our iterative algorithm converges again
after some iterations to a point at which all users achieve their
target SINRs. In addition, due to increased interference, all
users (users 1, 2, and 3) increase their transmit power levels and
decrease their data rates to compensate for the added interference
and achieve their target SINRs.

\begin{figure}[t]
  \begin{center}
    \includegraphics[width=9 cm , height=6 cm]{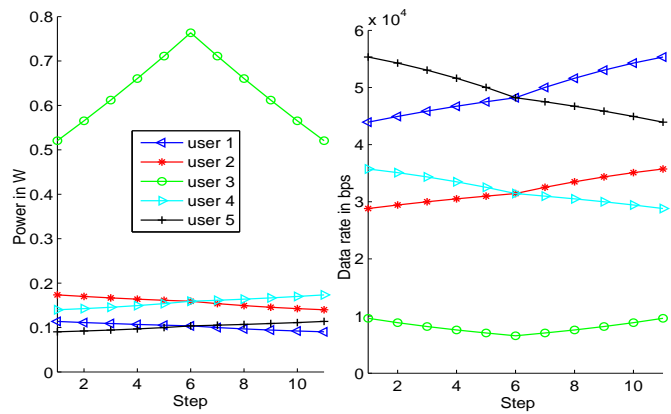} 
    \caption{Transmit power levels and data rates in our proposed resource allocation algorithm. Each step on the horizontal axis corresponds to a sufficient number of iterations needed for convergence.} 
    \label{fig:Power_rate_SINR_base station}
  \end{center}
\end{figure}

In our next simulation, we examine base station assignment by
considering two base stations and five users. All users are fixed
except user 3 that moves 10 meters away from base station 1
towards base station 2 at each step. Users' distances from base
stations 1 and 2 are $[110, 130, 210, 390, 410]$ and $[410, 390,
310, 130, 110]$, respectively at step 1, $[110, 130, 220, 390,
410]$ and $[410, 390, 300, 130, 110]$, respectively at step 2, and
$[110, 130, 310, 390, 410]$ and $[410, 390, 210, 130, 110]$,
respectively at the last step. The results are shown in Fig.
\ref{fig:Power_rate_SINR_base station}, where we show the
converged values at each step. From the first step to the sixth
step, the base station to which user 3 is assigned is base station
1. Note that in step 6, user 3 is located at equal distances from
the two base stations, i.e., 260 meters. However, when user 3 goes
further towards base station 2 in step 7, our scheme allocates it
to base station 2. In next steps, since user 3 goes near base
station 2, its channel becomes better and hence it consumes less
power and transmits at a higher data rate, as shown in Fig.
\ref{fig:Power_rate_SINR_base station}.

We now compare our proposed scheme to \cite{musku1} which is
recently published and is very close to our work. We first provide
results for two scenarios, where in the first one, users in the
network are situated at different distances from the base station,
and in the second, they have the same distance from the base
station. The results are shown in Table \ref{comparisontable}. In
Scenario 1, the network consists of one base station and 5 users
located at [110, 130, 210, 130, 150] meters from the base station.
We set the packet length to 100 bits in \cite{musku1},
$\alpha_1=10^6$, $\alpha_2=12.9492$, and $\lambda=0.0004$ for all
users. The value of $\lambda=0.0004$ results in achieving the
target SINR which is set by the chosen values of $\alpha_1$ and
$\alpha_2$. The value of $\alpha_2$ for two scenarios is set in
such a way that our scheme achieves the same SINR as in
\cite{musku1}. Besides, $P_i^\text{min}=10^{-6}$ Watts,
$P_i^\text{max}=0.1605$ Watts, $R_i^\text{min}=0.1$ bps, and
$R_i^\text{max}=96000$ bps. The value of $P_i^\text{max}$ is
chosen so that the maximum achieved transmit power would be the
same in both schemes. For Scenario 1, the total data rate of users
in \cite{musku1} is 110890 bps, while this value in our scheme is
99852 bps (slightly less than that in \cite{musku1}), but the
total transmit power in our scheme is 0.3914 watts, which is
significantly less than 0.8025 watts in \cite{musku1}. In other
words, we consume nearly 50\% less power at the cost of a slight
reduction in total throughput. In addition, one can see that the
user with the worst channel, i.e., user 3, attains a higher
throughput (i.e., a better degree of fairness) in our scheme than
in \cite{musku1}. In Scenario 2, we consider 5 users, each
situated 110 meters from the base station. The setting here is the
same as in Scenario 1, except for $P_i^\text{max}=0.0486$ Watts.
Note that in Table \ref{comparisontable} for Scenario 2, the
settings result in users' transmit power levels, users' data
rates, and users' attained SINRs to be the same in both schemes.

Next, we compare our proposed scheme with \cite{musku1} for
different number of users. The results are shown in Table
\ref{comparisontabledifferentusers}. Each row in the table
corresponds to one comparison. Note that we use the same settings
as in Scenario 2, i.e., $P_i^\text{max}=0.0647$ Watts for both
schemes, and transmit power, data rate, and achieved SINR are the
same for all users. The first 5 rows in Table
\ref{comparisontabledifferentusers} correspond to 5 different
cases, each with 3, 4, 5, 6, and 7 users all located at the same
distance of 110 meters from the base station. As can be seen in
Table \ref{comparisontabledifferentusers}, throughput values of
users in the first two rows are the same in both schemes. However,
for the same value of the total throughput for our scheme and
\cite{musku1}, power consumption in our scheme is less than
$P_i^\text{max}$, which is the transmit power in \cite{musku1}.
When the number of users in the network increases from 5 to 6 and
7, as the power level of users reaches its upper bound, the data
rates are obtained from (\ref{rateupdater}). Note that, although
the throughput of users in our scheme is higher than that in
\cite{musku1}, the achieved SINR values of users are less than
their target values as the convergence point of the algorithm is
out of the power strategy space. However, there exists one
additional degree of freedom in our scheme, i.e., the pricing
$\lambda$. In rows 6 and 7 of Table
\ref{comparisontabledifferentusers}, we again simulate the network
with 6 and 7 users but with $\lambda=0.0005$ and $\lambda=0.0006$,
respectively. Due to the higher pricing, users tend to transmit at
less power and/or at less data rate. For these values for the
pricing factor, the power level, the data rate, and the achieved
SINR of users are the same in both scheme.

Finally, we compare the performance of our proposed scheme with
that of \cite{musku1} when the distances of users change. The
network consists of 10 users and 1 base station. We set noise
power to $N_0=10^{-10}$, $\lambda=0.0001$, and $P_i^\text{max}=1$
Watts for all users. All other settings are the same as before
except for the distance of users. Users are located at the same
distance from the base station. The distance of users to base
station at each run is 50, 150, 250, and 350 meters, respectively.
The results are shown in Table
\ref{comparisontabledifferentdistance}. Note that when users are
located near to the base station, the data rates of all users are
equal in both schemes. However, our proposed scheme consumes much
less power than that of \cite{musku1}. When users are located at
350 meters from the base station, our scheme reaches its upper
bound on the power level. Similar to the previous cases in Table
\ref{comparisontabledifferentusers}, although the data rate of
users are higher in our scheme, the achieved SINR is less, meaning
that users did not achieve their target SINRs. However, we can
increase the pricing so that all users achieve their target SINRs.
This is shown in the last row of Table
\ref{comparisontabledifferentdistance}.

\section{Conclusions}\label{conclusion}
In this paper, we proposed a distributed scheme for joint data
rate and power control in CDMA networks. We formulated the problem
by utilizing a game theoretic framework, proved the existence and
uniqueness of NE for the game, and proved that our distributed
algorithm always converges to this unique NE. Also, we showed that
at NE, each user attains its target SINR, and that the proposed
algorithm is computationally efficient and its signalling overhead
is low. Simulation results validated our analysis.


\clearpage



\clearpage

\begin{table}
\caption{An Example in Which the Convergence Point Is Out of
Strategy Space, Where the Effects of Increased Pricing and User
Removal on the Convergence Point Are Shown} \vspace{-0.2in}


\begin{tabular}{||@{ }c@{ }||c|c|c||c|c|c||c|c|c||}
  \hline \hline
    \multirow{2}[4]{.5 cm} & \multicolumn{3}{c||}{$\lambda=10^{-5}$ } &
  \multicolumn{3}{c||}{$\lambda=10^{-4}$} & \multicolumn{3}{c||}{User 3 is Removed}
  \\\cline{2-10}
    & $p_i$ in W   & $r_i$ in bps & Target SINR & $p_i$ in W & $r_i$ in bps & Target SINR & $p_i$ in W & $r_i$ in bps & Target SINR
   \\\hline \hline
    User 1 & 1.011 & 47000 & 21.0452 & 0.1127 & 44360 & 20 & 0.08     & 47000 &  26.58 \\ \hline
   User 2 & 1.5533 & 32189 & 20      & 0.172  & 29075 & 20 &  0.125   & 40000 & 20 \\ \hline
   User 3 & 3      & 10205 & 12.2458 & 0.5166 & 9679  & 20  & -    & -     & - \\
   \hline \hline
\end{tabular}
   \label{tableboundingstrategy}
\end{table}

\begin{table}
\caption{Power Level, Data Rate, and the Achieved SINR for Each
User in Our Proposed Scheme and in \cite{musku1} When Users Are
Located at Different Distances (Scenario 1) and at Equal Distance
(Scenario 2) from the Base Station} \vspace{-0.2in} \centering
\begin{tabular}{||c|c||c|c|c||c|c|c||}
  \hline \hline
  \multicolumn{2}{||l||}{\multirow{2}[4]{1.4cm}} & \multicolumn{3}{c||}{Our Proposed Scheme} & \multicolumn{3}{c||}{The Scheme in \cite{musku1}}
  \\\cline{3-8}
    \multicolumn{2}{||l||}{} & $p_i$ in W& $r_i$ in bps & Target SINR & $p_i$ in W & $r_i$ in bps & Target SINR
   \\\hline \hline
      \multirow{5}[4]{1.4cm}{Scenario 1}  & User 1 & 0.0388 & 32201 & 12.9492 & 0.1605 & 55567 & 12.9492  \\ \cline{2-8}
  & User 2 & 0.0569 & 21949 & 12.9492 & 0.1605 & 21089 & 12.9492 \\ \cline{2-8}
   & User 3 & 0.1605 & 7787 & 12.9492 & 0.1605 & 2511 & 12.9492 \\ \cline{2-8}
    & User 4 & 0.0569 & 21949 & 12.9492 & 0.1605 & 21089 & 12.9492 \\ \cline{2-8}
  & User 5 & 0.0782 & 15982 & 12.9492 & 0.1605 & 10632 & 12.9492 \\
  \hline \hline
  \multirow{5}[4]{1.4cm}{Scenario 2}  & User 1 & 0.0647 & 19306 & 12.9492 & 0.0647 & 19306 & 12.9492 \\ \cline{2-8}
   & User 2 & 0.0647 & 19306 & 12.9492 & 0.0647 & 19306 & 12.9492 \\ \cline{2-8}
   & User 3 & 0.0647 & 19306 & 12.9492 & 0.0647 & 19306 & 12.9492 \\ \cline{2-8}
    & User 4 & 0.0647 & 19306 & 12.9492 & 0.0647 & 19306 & 12.9492 \\ \cline{2-8}
   & User 5 & 0.0647 & 19306 & 12.9492 & 0.0647 & 19306 & 12.9492 \\
   \hline \hline
\end{tabular}
\label{comparisontable}
\end{table}

\begin{table}
\caption{Comparison of the Performances of Our Proposed Scheme
with That of \cite{musku1} for Different Number of Users and
Pricing $\lambda$} \vspace{-0.2in} \centering
\begin{tabular}{||c|c||c|c|c||c|c|c||}
  \hline \hline
  \multirow{2}[3]{1.4cm}{Number of Users} & \multirow{2}[3]{1.4cm}{Pricing Factor} & \multicolumn{3}{c||}{Our Proposed Scheme} & \multicolumn{3}{c||}{The Scheme in \cite{musku1}}
  \\\cline{3-8}
    &  & $p_i$ in W& $r_i$ in bps & Target SINR & $p_i$ in W & $r_i$ in bps & Target SINR
   \\\hline \hline
      M=3  & $\lambda=4 \times 10^{-4}$ & 0.0324 & 38612 & 12.9492 & 0.0647 & 38612 & 12.9492 \\ \cline{1-8}
   M=4  & $\lambda=4 \times 10^{-4}$ & 0.0486 & 25741 & 12.9492 & 0.0647 & 25741 & 12.9492 \\  \cline{1-8}
    M=5  & $\lambda=4 \times 10^{-4}$  & 0.0647 & 19306 & 12.9492 & 0.0647 & 19306 & 12.9492 \\ \cline{1-8}
     M= 6 & $\lambda=4 \times 10^{-4}$  & 0.0647 & 17274 & 11.578 & 0.0647 & 15445 & 12.9492 \\ \cline{1-8}
      M=7  & $\lambda=4 \times 10^{-4}$  & 0.0647 & 15769 & 10.569 & 0.0647 & 12871 & 12.9492 \\
  \hline \hline
       M= 6 & $\lambda=5 \times 10^{-4}$  & 0.0647 & 15445 & 12.9492 & 0.0647 & 15445 & 12.9492 \\ \cline{1-8}
      M=7  & $\lambda=6 \times 10^{-4}$  &  0.0647 & 12871 & 12.9492 & 0.0647 & 12871 & 12.9492 \\
  \hline \hline
\end{tabular}
\label{comparisontabledifferentusers}
\end{table}

\begin{table}
\caption{Comparison of the Performances of Our Proposed Scheme
with That of \cite{musku1} for Different Distances of Users and
Pricing $\lambda$} \vspace{-0.2in} \centering
\begin{tabular}{||c|c||c|c|c||c|c|c||}
  \hline \hline
  \multirow{2}[3]{1.4cm}{User Distance} & \multirow{2}[3]{1.4cm}{Pricing Factor} & \multicolumn{3}{c||}{Our Proposed Scheme} & \multicolumn{3}{c||}{The Scheme in \cite{musku1}}
  \\\cline{3-8}
    &  & $p_i$ in W& $r_i$ in bps & Target SINR & $p_i$ in W & $r_i$ in bps & Target SINR
   \\\hline \hline
      $d=50$ Meters  & $\lambda=1 \times 10^{-4}$ & 0.583 & 8570 & 12.9492 & 1 & 8570 & 12.9492 \\ \cline{1-8}
   $d=150$ Meters & $\lambda=1 \times 10^{-4}$ & 0.635 & 8110 & 12.9492 & 1 & 8110 & 12.9492 \\  \cline{1-8}
    $d=250$ Meters & $\lambda=1 \times 10^{-4}$  & 0.879 & 5686 & 12.9492 & 1 & 5686 & 12.9492 \\ \cline{1-8}
     $d=350$ Meters & $\lambda=1 \times 10^{-4}$  & 1 & 3972 & 10.287 & 1 & 3155 & 12.9492 \\ \cline{1-8}
      \hline \hline
      $d=350$ Meters & $\lambda=1.6 \times 10^{-4}$  &  1 & 3155 & 12.9492 &  1 & 3155 & 12.9492 \\ \cline{1-8}
       \hline \hline
\end{tabular}
\label{comparisontabledifferentdistance}
\end{table}

\end{document}